\documentclass[sigconf,screen]{acmart}

\usepackage{00_preamble}
\usepackage{00_macros}

\begin{document}

\begin{abstract}
Multi-agent collaboration enhances situational awareness in intelligence, surveillance, and reconnaissance (ISR) missions. Ad hoc networks of unmanned aerial vehicles (UAVs) allow for real-time data sharing, but they face security challenges due to their decentralized nature, making them vulnerable to cyber-physical attacks. This paper introduces a trust-based framework for assured sensor fusion in distributed multi-agent networks, utilizing a hidden Markov model (HMM)-based approach to estimate the trustworthiness of agents and their provided information in a decentralized fashion. Trust-informed data fusion prioritizes fusing data from reliable sources, enhancing resilience and accuracy in contested environments. To evaluate the assured sensor fusion under attacks on system/mission sensing, we present a novel multi-agent aerial dataset built from the Unreal Engine simulator. We demonstrate through case studies improved ISR performance and an ability to detect malicious actors in adversarial settings.
\end{abstract}

\title{Trust-Based Assured Sensor Fusion in Distributed \protect\\ Aerial Autonomy}
\date{}

\newif\ifAnonymize

\Anonymizefalse

\ifAnonymize

\else
    \author{R. Spencer Hallyburton}
    \affiliation{%
    \institution{Department of Electrical and Computer Engineering}
      \institution{Duke University}
      \city{Durham}
      \state{NC}
      \country{USA}}
    \email{spencer.hallyburton@duke.edu}
    \author{Miroslav Pajic}
    \affiliation{%
    \institution{Department of Electrical and Computer Engineering}
      \institution{Duke University}
      \city{Durham}
      \state{NC}
      \country{USA}}
    \email{miroslav.pajic@duke.edu}
    \setcopyright{acmlicensed}
    \copyrightyear{2025}
    \acmYear{2025}
    \setcopyright{cc}
    \setcctype{by}
    \acmConference[ICCPS '25]{ACM/IEEE 16th International Conference on Cyber-Physical Systems (with CPS-IoT Week 2025)}{May 6--9, 2025}{Irvine, CA, USA}
    \acmBooktitle{ACM/IEEE 16th International Conference on Cyber-Physical Systems (with CPS-IoT Week 2025) (ICCPS '25), May 6--9, 2025, Irvine, CA, USA}
    \acmDOI{10.1145/3716550.3722038}
    \acmISBN{979-8-4007-1498-6/2025/05}
\fi



\maketitle
 
\section{Introduction}

Collaboration among autonomous vehicles (AVs), such as unmanned aerial vehicles (UAVs), significantly enhances situational awareness for intelligence, surveillance, and reconnaissance (ISR), where timely and accurate information is crucial. UAVs are attractive for dynamic environments due to their low cost and mobility~\cite{bendea2008low}. A collection of networked UAVs can execute collaborative ISR forming an ad hoc sensor network. While each agent performs local perception and navigation, broad situational awareness is obtained by sharing compact data payloads over communication channels and fusing data with classical distributed data fusion (DDF) algorithms~\cite{1994ddfframework}.

Unfortunately, data fusion in ad hoc sensor networks faces significant security challenges. Unlike traditional networks with secure central processing for data aggregation, ad hoc networks lack fixed infrastructure, instead relying on each node to relay data. This distributed nature makes sensor networks susceptible to attacks such as eavesdropping, data tampering, node capture, and false data injection~\cite{hu2005security,yaacoub2020security}. Thus, ad hoc networks require adaptive, lightweight security protocols to maintain resilience while balancing~efficiency.

Security of cyber-physical systems (CPS) requires both \emph{proactive} and \emph{reactive} threat mitigation~\cite{kwon2014proactive,yue2007intrusion}. Compromised nodes in sensor networks present unique challenges to proactive security because e.g.,~an adversarial UAV operates as a network insider, bypassing cryptographic protections. As a result, \emph{reactive} security is necessary to ensure reliable data fusion. Existing reactive protocols employ Byzantine fault models to detect threats and eliminate compromised agents~\cite{kihlstrom2003byzantine}. However, these models fail to address the specific challenges of CPS, particularly for UAVs, where natural false positives/negatives, dynamic sensor occlusions, and sensor degradation are common~\cite{grigoropoulos2020byzantine}. These natural issues can mimic or obscure actual threats, making traditional Byzantine approaches unsuitable for real-world autonomy. To meet these unique challenges, a more flexible and adaptive security model is needed.

Consequently, to address the gap in reactive security for distributed aerial autonomy, we propose a two-part framework for \emph{trust-based assured sensor fusion}. Trust measures the consistency of provided information with physical laws via kinematic models and information from other trusted agents, and it is used to identify and filter distrusted entities. Part (1) of assured fusion estimates probability density functions (PDFs) of the trustedness of agents and their provided information. Part (2) leverages trust estimates to weight each agent's influence in data fusion. The assured sensor fusion pipeline both detects misbehaving agents and recovers accurate situational awareness in the face of adversarial data manipulation.

In a distributed context, each agent performs its own local trust estimation process using data shared among nearby agents. Trust estimates are represented by parametric Beta distributions over the domain $[0,\,1]$, as we previously proposed in~\cite{hallyburton2024bayesian}, and we frame trust estimation as hidden-state estimation on a three-step hidden Markov model~(HMM). In the first step, each ``ego'' agent propagates the previous trust estimates to the current time. In the second step, the ego agent evaluates data shared by nearby ``proximal'' agents to yield trust measurements. Finally, trust measurements update trust PDF estimates of all connected agents and their tracked objects.

Since there is no sensor providing direct measurements of trust, the ego agent assesses the consistency of data from proximal agents to derive \emph{pseudomeasurements} (PSMs) of trust. The PSMs, expressed as a tuple of value and confidence on $[0,\,1]$, are generated with pairwise comparisons between each agent's most recent situational awareness and a prediction of what the agent \emph{should} have seen based on its position, sensing capability, and field of view (FOV). The PSMs iteratively update distributions for \emph{agent trust} and \emph{track trust} in a process inspired by Gibbs sampling~\cite{shemyakin2017copula}.

With platform-local trust estimates in hand, we propose a novel trust-weighted DDF algorithm for fusing data received from proximal agents. Our trust-informed data fusion algorithm prioritizes information from trusted agents, building resiliency to identified malicious actors. Each piece of data provided to the ego by a proximal agent is weighted by the expectation over the estimated trust distribution of that agent from the ego's local trust estimation process. To represent the degree of uncertainty of fusing data from multiple agents of varied trust, we derive a trust-based fusion confidence measure. We furthermore construct safeguards to handle edge cases including minimal FOV overlap between agents with trust initialization based on a-priori agent trustedness. 

Evaluating assured distributed sensor fusion in aerial autonomy requires multi-agent UAV datasets. Leveraging the dataset generation pipeline from~\cite{hallyburton2023datasets}, we construct the first multi-agent aerial datasets using CARLA simulator~\cite{dosovitskiy2017carla}. We deploy multiple UAVs with sensing capability within a simulated environment, capturing high-rate data and performing our own ground truth labeling. From this baseline dataset of UAV missions, we apply randomization to sensor noise, communication capability, and agent activity to create a corpus of datasets for Monte Carlo evaluation. Finally, we apply randomized adversary models for large scale evaluations of trust in contested aerial autonomy. The dataset is released open source~\cite{trust-links}.

Classical metrics on sensor fusion and novel metrics on trust illustrate the effectiveness of trust-based assured sensor fusion in mitigating threats from adversaries. We motivate the need for networked UAVs by demonstrating the positive influence of increased \emph{density} of agents on trust observability and system resilience. Moreover, we present large scale evaluations highlighting the importance of accurate prior information and illustrating the influence of varied attacker capability to demonstrate that trust-based assured sensor fusion is a promising solution for assured autonomy.

\vspace{4pt}
\noindent\textbf{Contributions.} \ In summary, the main contributions of this work~are:
\begin{itemize}[leftmargin=12pt,topsep=0pt]
    \item A framework for distributed multi-agent data fusion for ISR in ad hoc networks of autonomous aerial agents.
    \item The first dataset of perception and navigation data for multi-agent aerial autonomy using the CARLA simulator.
    \item An HMM-based methodology for trust estimation of agents and tracked objects in ad hoc aerial networks.
    \item Novel trust-aware distributed data fusion that integrates trust distributions into classical fusion.
    \item Demonstration of the assured sensor fusion on case studies and Monte Carlo evaluations of UAVs in contested environments.
\end{itemize}

\vspace{4pt}
\noindent This paper is organized as follows. Section~\ref{sec:background} describes shortcomings of prior approaches to trust-based sensor fusion. Section~\ref{sec:system-model} illuminates models of the platforms and algorithms in unmanned aerial autonomy; the vulnerability of such systems follows in Section~\ref{sec:threat-model}. We present our novel framework for trust estimation and trust-informed data fusion in Section~\ref{sec:trust-model} before finally illustrating its effectiveness in Sections~\ref{sec:multi-agent-sim} and~\ref{sec:experiments} with case studies and evaluations.
\section{Background} \label{sec:background}

\subsection{Distributed Data Fusion (DDF)}

Ad hoc sensor networks improve coverage, resilience, and response times in dynamic environments without fixed infrastructure. However, due to its distributed nature, ad hoc data fusion may suffer from unknown and mis-modeled correlations between platforms. Ad hoc networks can also suffer from rumor propagation where data traverses the network and is double-counted~\cite{1994ddfframework}. This may lead to overconfidence of situational awareness gathered from distributed platforms. DDF algorithms counter inter-agent correlations in distributed applications by performing conservative state estimation such as with the covariance intersection (CI) algorithm~\cite{julier2017general}.

\subsection{Trust in Wireless Sensor Networks}

Trust has long been considered important for wireless sensor networks (WSNs); a survey of classical frameworks including the first proposals for dynamic, recursive trust estimation~\cite{liu2004dynamic,zhu2004computing} can be found in~\cite{yu2012trust}. Unfortunately, few of such works consider the practical implications of receiving noisy sensor data in dynamic environments. Moreover, mobility and FOV restrictions are significant barriers to adopting these classical approaches. Contemporary models including~\cite{cavorsi2024exploiting,cheng2021trust} provide frameworks for trust-informed decision in mobile platforms, but yield no guidance regarding generation of trust measurements from real sensor data (including perception-based sensing); trust measurements are assumed provided by an oracle. Moreover, few contemporary works consider the impact of uncertainty or how to incorporate available prior information, significantly limiting the robustness of trust-based sensor fusion. 

\subsection{Misbehavior Detectors (MBDs)}

In contested environments, since it is not known which agents are compromised, security awareness is important. Misbehavior detectors (MBDs) are algorithms that detect abnormal nodes such as those sending false data or disrupting communications~\cite{van2018survey}. Such algorithms use statistical and machine-learning-based methods for monitoring inconsistencies~\cite{golle2004detecting,bissmeyer2012assessment}. Unfortunately, MBDs suffer from challenges in sensor networks with unknown topologies. Moreover, many MBDs are isolated integrity functions that are not integrated in the sensor fusion process and are thus suboptimal in obtaining assured distributed and collaborative situational awareness.

\subsection{Trust-Based Methods in Aerial Autonomy}

Some multi-agent, multiple-target tracking (MTT) algorithms have features that promote data integrity. MTT algorithms employ recursive likelihood ratio tests to build track scores denoting the probability that a track represents a true object~\cite{1986blackmanRadar}. While track scoring handles natural false positive/negative, it was proven to be vulnerable to adversarial manipulation~\cite{hallyburton2024bayesian}. Existing security-aware sensor fusion is limited in handling real-world cases with time-varying trust, sensor uncertainties, and occlusions~\cite{golle2004detecting,huhns2002trusted}. Recent works considered trust of ground-vehicles in urban environments~\cite{hallyburton2025fusionmate}. 
\section{Collaborative Multi-Agent Autonomy} \label{sec:system-model}

In this section, we describe characteristics of collaborative multi-agent autonomy in aerial applications. We then present a framework for distributed data fusion and discuss relevant algorithms including perception, tracking, inter-platform communication, and data~fusion.

\begin{figure}[!t]
    \centering
    \fbox{\includegraphics[trim={0 1cm 0 2cm},clip,width=0.7\linewidth]{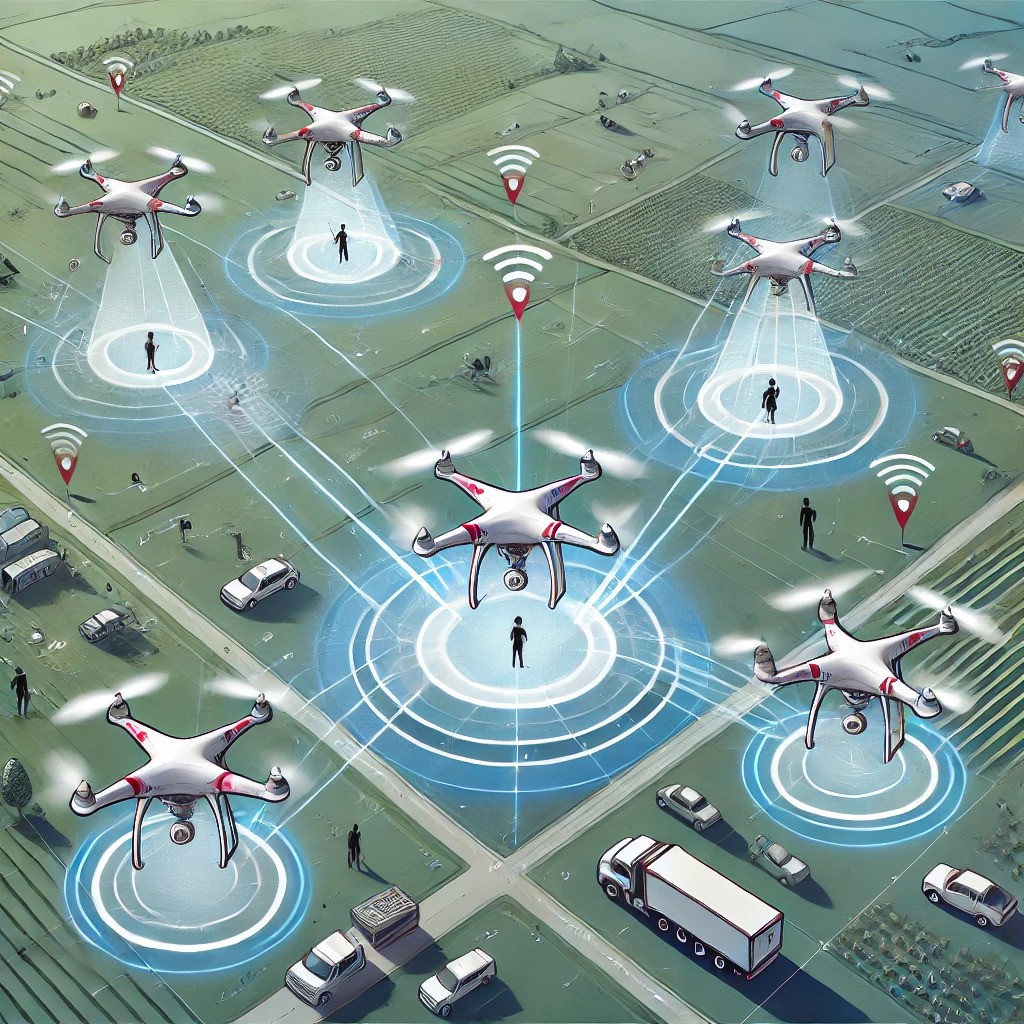}}
    \caption{Multiple aerial vehicles collaborate to detect and track objects. UAVs are individually responsible for small areas and tracks are shared among connected agents. Local to each platform are algorithms for detection/tracking platform-local data. DDF on each platform ingests data from nearby agents to create synthesized operating picture.}
    \label{fig:aerial-surveillance}
\end{figure}

\subsection{Collaborative Mission Statement}

Collaborative ISR over large areas can be achieved with networks of UAVs. The proliferation of wireless technology such as WiFi/5G has even enabled collaboration between multiple \emph{providers} of platforms (e.g.,~international collaboration). We consider that multiple aerial agents are equipped with a downward facing (bird's eye view, BEV), wide angle camera that detects and tracks ground-level objects using local algorithms, as illustrated in Figure~\ref{fig:aerial-surveillance}). 

\begin{figure}[!t]
    \centering\includegraphics[width=0.96\linewidth]{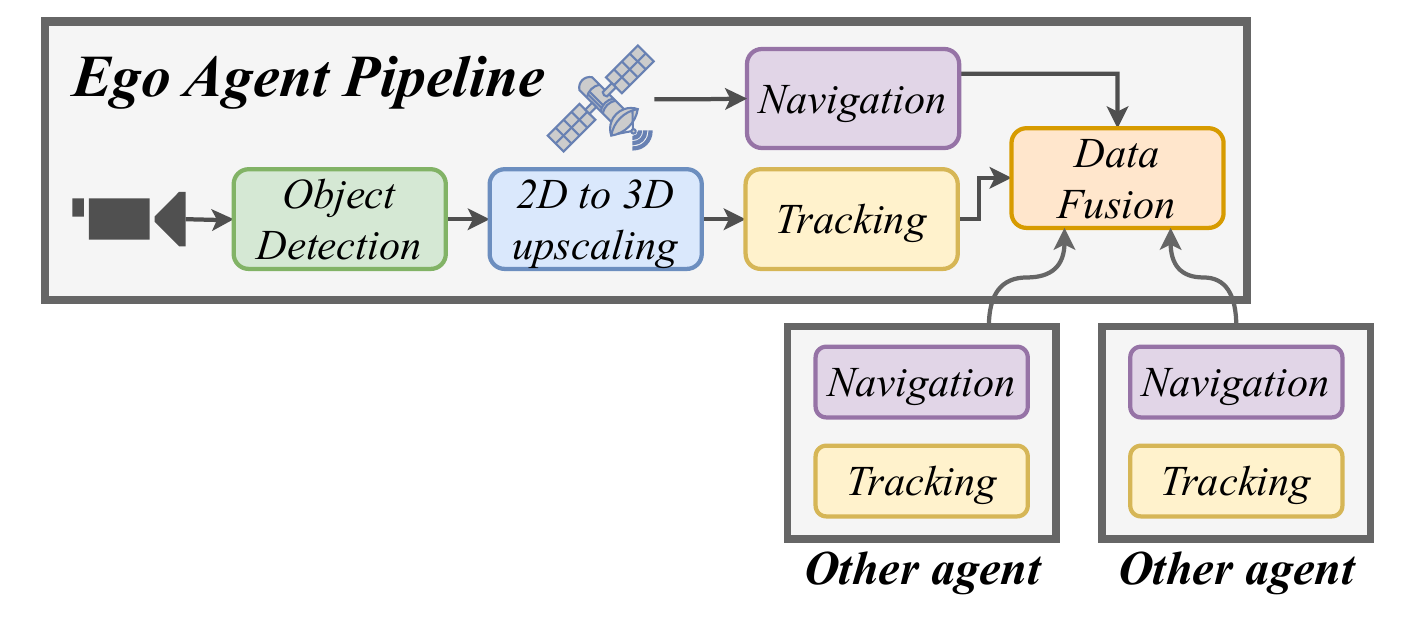}
    \vspace{-10pt}
    \caption{Ego agent performs local computations for detection and tracking of objects and navigation from platform sensing data. Ego communicates with nearby agents and executes conservative data fusion over the received object data.}
    \label{fig:multi-agent-fusion}
\end{figure}

Each agent's field of view (FOV) is a small subset of the global environment. Agents share data over communication channels for enhanced situational awareness, as in Figure~\ref{fig:multi-agent-fusion}. In the next sections, we describe the data fusion algorithms that enable collaborative ISR. The complete DDF pipeline is described in Algorithm~\ref{alg:ddf-aerial}. 

\begin{algorithm}
\caption{Distributed Data Fusion in Aerial Autonomy}
\label{alg:ddf-aerial}
\begin{algorithmic}[1]
\Require Ego agent image $I_t$, previous timestep ego tracks $\mathcal{T}_{t-1}$, proximal agent poses $\mathcal{A} = \{\mathcal{A}^k_t, \ k=1, ..., K\}$
\State Detect objects: $\mathcal{D}_{t} = \texttt{detect}(I_{t})$
\State Track objects: $\mathcal{T}_{t} = \texttt{track}(\mathcal{D}_{t}, \mathcal{T}_{t-1})$
\State Receive tracks from $\hat{K}$ nearby agents:
\[
\mathcal{T}^{\hat{K}}_t = \{ \mathcal{T}^k_t \text{ if } \texttt{communicate}(\mathcal{A}^{\text{ego}}_t,\ \mathcal{A}^k_t), k=1,...,K \}
\]
\State Distributed data fusion: $\mathcal{T}_{t,\text{fused}} = \texttt{DDF}(\mathcal{T}_t,\  \mathcal{T}^{\hat{K}}_t)$
\State \Return $\mathcal{T}_{t,\text{fused}}$
\end{algorithmic}
\end{algorithm}


\subsection{Algorithms for Platform-Local Data}

Agents execute algorithms to detect and track objects from downward facing camera data. We now describe the typical suite of algorithms employed for such tasks.

\subsubsection{Perception.} \ A perception model based on deep neural networks (DNNs) runs inference to detect objects from camera images. Camera detections are two-dimensional bounding boxes in pixel coordinates within the image.

\subsubsection{2D to 3D box upscaling.} 
\begin{figure}[!t]
    \centering
    \includegraphics[width=\linewidth]{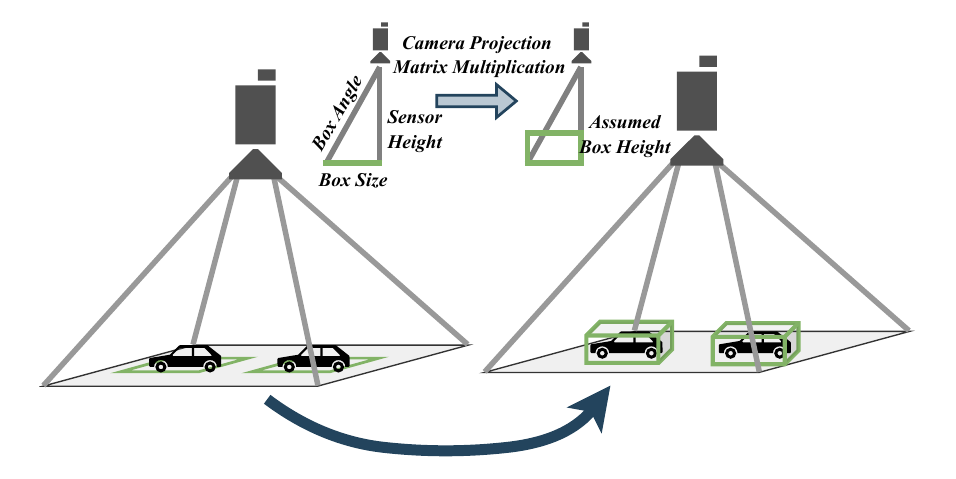}
    \vspace{-10pt}
    \caption{Ground constraint on detected object location allows upscaling 2D image detections to 3D boxes in Cartesian coordinates despite camera not directly measuring range.}
    \label{fig:box-upscale}
\end{figure}

\ With accurate ownship localization and knowledge of the camera's calibration, assuming detected objects are \emph{ground objects} enables upscaling the pixel-space box detections to fully-specified bounding boxes in Cartesian coordinates. Figure~\ref{fig:box-upscale} illustrates the upscaling process.
 
\subsubsection{Tracking.} \ Detections are tracked locally with a standard ten-state Kalman filter. The state vector is composed of position $\left[p_x, \, p_y, \, p_z\right]$, velocity $\left[v_x, \, v_y, \, v_z\right]$, box size $\left[h, \, w, \, l\right]$, and box orientation $\theta$ parameters,
\begin{align*}
    \hat{x} = \left[p_x, \, p_y, \, p_z, \, v_x, \, v_y, \, v_z, \, h, \, w, \, l, \, \theta \right].
\end{align*}
Objects are propagated forward in time using a constant velocity motion model and compensating for any platform motion. Data association between existing tracks and incoming detections uses the intersection over union (IoU) affinity metric.

\subsubsection{Field of view estimation.} \ Data fusion accuracy is enhanced when the sensor field of view (FOV) is known. The FOV aids in predicting the set of objects that platforms should have observed. We assume that the BEV camera is gimbaled and that two sensing axes remain parallel to the ground plane. The FOV is thus consistent and unimpeded between the platform and the ground. This is in contrast to e.g., ground vehicle sensors that suffer from strong occlusions in dense urban environments and require FOV estimation algorithms such as ray tracing or segmentation~\cite{hallyburton2025fov}.

\subsection{Inter-Platform Collaboration}

Sharing data between platforms requires modeling the communication capability, defining the information to be shared, and describing a strategy for fusing shared information.

\paragraph{Natural adversity.} \ Natural challenges to data fusion include latency and communication drops. We neglect issues from latency/drops since these can be handled by maintaining TCP data buffers.

\subsubsection{Communication.} \ We assume agents have an omnidirectional antenna for communication. Bandwidth is limited, so agents judiciously send only the following vectors/matrices:
\begin{itemize}
    \item Ownship estimated position vector/covariance;
    \item Camera position/orientation on platform (fixed);
    \item Camera intrinsic parameters (fixed);
    \item Each track's estimated state vector/covariance.
\end{itemize}

Each ownship position is six dimensional and composed of three position and three orientation states (42 floats). The camera's FOV function is defined by the platform's position, the camera position and orientation relative to the platform (6 floats), and the camera intrinsic parameters of horizontal and vertical camera fields of view, focal length, and horizontal and vertical image sizes (3 floats, 2 ints). The track state estimates are seven dimensional (56 floats). Thus, each packet is composed of 107 floats and 3 ints totaling 3448 bits. At 10~Hz transmission, this amounts to just under 35~Kbps - a low data rate requiring very little bandwidth. On the other hand, transmitting a three-channel full-HD image at 10~Hz requires nearly 500~Mbps without compression techniques.


\subsubsection{Fusion.} \ Formally, the multi-agent sensor fusion problem can be captured as estimating the object state posterior
\begin{equation} \label{eq:fusion-posterior}
\begin{aligned}
    \Pr(X_{t} | Z_{1:t}, A_{1:t}),
\end{aligned}
\end{equation}
where $X_{t}$ are states for all $i=1...N$ objects, $Z_{1:t}$ are data from all $k=1...K$ agents, and $A_{1:t}$ are agent characteristics including pose and sensor models. The steps for data fusion are as follows.

\begin{figure}
    \centering
    \includegraphics[width=0.96\linewidth]{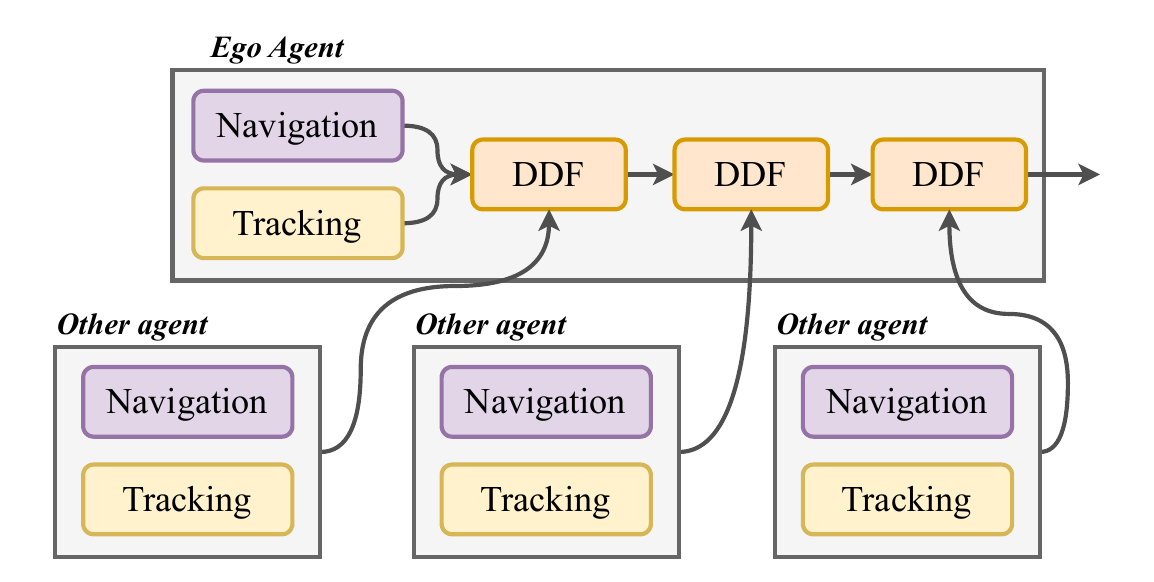}
    \caption{Cascaded, pairwise distributed data fusion (DDF) scales classical DDF to multiple connected agents.}
    \label{fig:cascaded-ddf}
\end{figure}

\paragraph{Frame registration.} \ Each proximal agent transmits tracks in its local reference frame. Tracks must be aligned with the ego agent's reference. Assuming a gimbaled sensor that maintains sensing axes parallel with the ground plane, an affine transformation suffices.

\paragraph{Clustering/assignment.} \ With reference frames aligned, a cost-minimizing assignment matches tracks between proximal and ego agents in regions of FOV overlap.

\paragraph{Data fusion.} \ In an ad-hoc network, distributed data fusion (DDF) handles fusing tracked objects between agents~\cite{1994ddfframework}. A cascaded approach is commonly used when fusing data from more than two agents. As in Figure~\ref{fig:cascaded-ddf}, the set of fused tracks begins as solely the ego's local tracks. Incrementally, tracks from each agent are fused with the existing set of fused tracks. Any tracks from the new agent not present in the fused set start new tracks in the fused set. We use the covariance intersection (CI) algorithm~\cite{1994ddfframework} to perform conservative fusion of any matched tracks between the fused set and the new agent. The pairwise CI algorithm is presented in Algorithm~\ref{alg:covariance-intersection}.

\begin{algorithm}
\caption{Pairwise Covariance Intersection (CI) Algorithm}
\label{alg:covariance-intersection}
\begin{algorithmic}[1]
\Require Covariance matrices $\Sigma_1$, $\Sigma_2$ and mean estimates $\mu_1$, $\mu_2$
\State Initialize weight parameter $\omega \in [0, 1]$
\While{not converged}
    \State Compute fused covariance: 
    \[
    \Sigma_{CI} = \left( \omega \Sigma_1^{-1} + (1 - \omega) \Sigma_2^{-1} \right)^{-1}
    \]
    \State Compute fused mean: 
    \[
    \mu_{CI} = \Sigma_{CI} \left( \omega \Sigma_1^{-1} \mu_1 + (1 - \omega) \Sigma_2^{-1} \mu_2 \right)
    \]
    \State Adjust $\omega$ to minimize a criterion (e.g., determinant of $\Sigma_{CI}$)
\EndWhile
\State \Return $\mu_{CI}, \Sigma_{CI}$
\end{algorithmic}
\end{algorithm}

\section{Threats Against Collaborative Sensing} \label{sec:threat-model}

Attackers can disrupt collaborative fusion by exploiting physical channels with spoofing attacks~\cite{2019cao-spoofing, 2022hally-frustum} or acting as network insiders with attacks at the data level, thus bypassing proactive security measures~\cite{hallyburton2023partial, 2014cyberattacks}; moreover, they can employ attacks over the communication channels if the network is not sufficiently secured~\cite{ansari2021v2x, monteuuis2018my}. 

\subsection{Attacker Goals}
We consider an attacker that disrupts situational awareness by compromising sensor and/or communicated data in a distributed context. By acting over physical, cyber, or communication channels, the attacker may introduce the following errors into sensor fusion:

\begin{itemize}
    \item \textbf{False positive:} Insert fake object(s) into detections.
    \item \textbf{False negative:} Remove true object(s) from detections.
    \item \textbf{Translation:} Translate detections spatially.
\end{itemize}

\subsection{Attacker Capability}
An attacker can compromise some number of agents in the environment. We consider cases where an attacker only compromises an isolated agent as well as when a subset of agents from e.g.,~the same manufacturer are compromised. We perform systematic evaluations over parametric attacker capabilities.

\subsection{Special Considerations for Autonomy}
In CPS, where agents are interacting in a dynamic environment, natural adversities/uncertainties may occur. These include natural false positives/negatives, moments of unexpected sensor occlusions, and adverse weather conditions. This is in contrast to fault-tolerant computing, where such ``natural'' adversities rarely occur.
\section{Assured Trust-Informed Sensor Fusion} \label{sec:trust-model}

A primary goal of multi-agent distributed data fusion is broad situational awareness; however, adversarial actors can corrupt this pipeline. For assured sensor fusion, agents in ISR missions must achieve a high degree of \emph{trustedness} to contribute to the network. This section defines trustedness in autonomy, introduces trust estimation algorithms in ad hoc sensor networks, and derives trust-informed DDF algorithms for assured fusion.

\subsection{Defining Trust in Autonomy}

We address the challenge of estimating the \emph{trust} of agents and tracks within a UAV network. Here, we consider \emph{trust} as an abstract measure of an entity’s \emph{trustworthiness} (see Definition~\ref{def:trust}), reflecting its competency and integrity in providing accurate data~\cite{huhns2002trusted}. Trust spans a spectrum from \emph{trusted} to \emph{distrusted}, with an intermediate state, \emph{untrusted}, indicating that the entity is neither fully trusted nor distrusted.

\begin{definition}[Trust]\label{def:trust}
    Trust indicates the extent to which information provided by an agent or about an object aligns with physical laws and accurately represents the true state of the environment.
\end{definition}

Quantifying trust, as in Definition~\ref{def:trust-est}, is valuable for identifying untrustworthy agents, preventing the spread of misinformation, and using trust to inform sensor fusion objectives. Monitoring \emph{agent trust} aids in identifying the \emph{sources} of misinformation, while monitoring \emph{track trust} helps in avoiding misguided \emph{actions} that may arise from maintaining incorrect situational awareness.

\begin{definition}[Trust Estimation] \label{def:trust-est}
    Trust estimation is a mapping of sensor data and agent attributes to the interval $[0,1]$, where 0 represents complete distrust and 1 represents complete trust.
\end{definition}

\begin{subdefinition}[Agent Trust, $\{\tau^a_k\} \coloneqq \Tau^a$ for agents $k = 1, \dots, K$]
    Agent trust refers to the degree to which information provided by an agent aligns with trusted agents and adheres to physical laws.
\end{subdefinition}

\begin{subdefinition}[Track Trust, $\{\tau^c_j\} \coloneqq \Tau^c$ for tracks $j = 1, \dots, J$]
    Track trust denotes the degree of confidence in a track’s existence and the consistency of its estimated state.
\end{subdefinition}

\paragraph{Trust as enhancement to Byzantine robustness.} \ The use of trust in autonomy requires distinct algorithms beyond the Byzantine fault-tolerance model. Unlike the classical approach where agents are excluded for deviating from consensus, AVs must account for natural errors such as false positives and negatives due to noisy, occluded sensor data. Moreover, trust should build gradually with consistent behavior, but degrade quickly with inconsistencies. Thus, UAV data fusion needs specialized trust frameworks—to the best of our knowledge, none existed before this work.

\subsection{Trust-Based Fusion with Bayesian Principles} \label{sec:security-aware-bayesian}

\begin{figure}
    \centering
    \includegraphics[width=\linewidth]{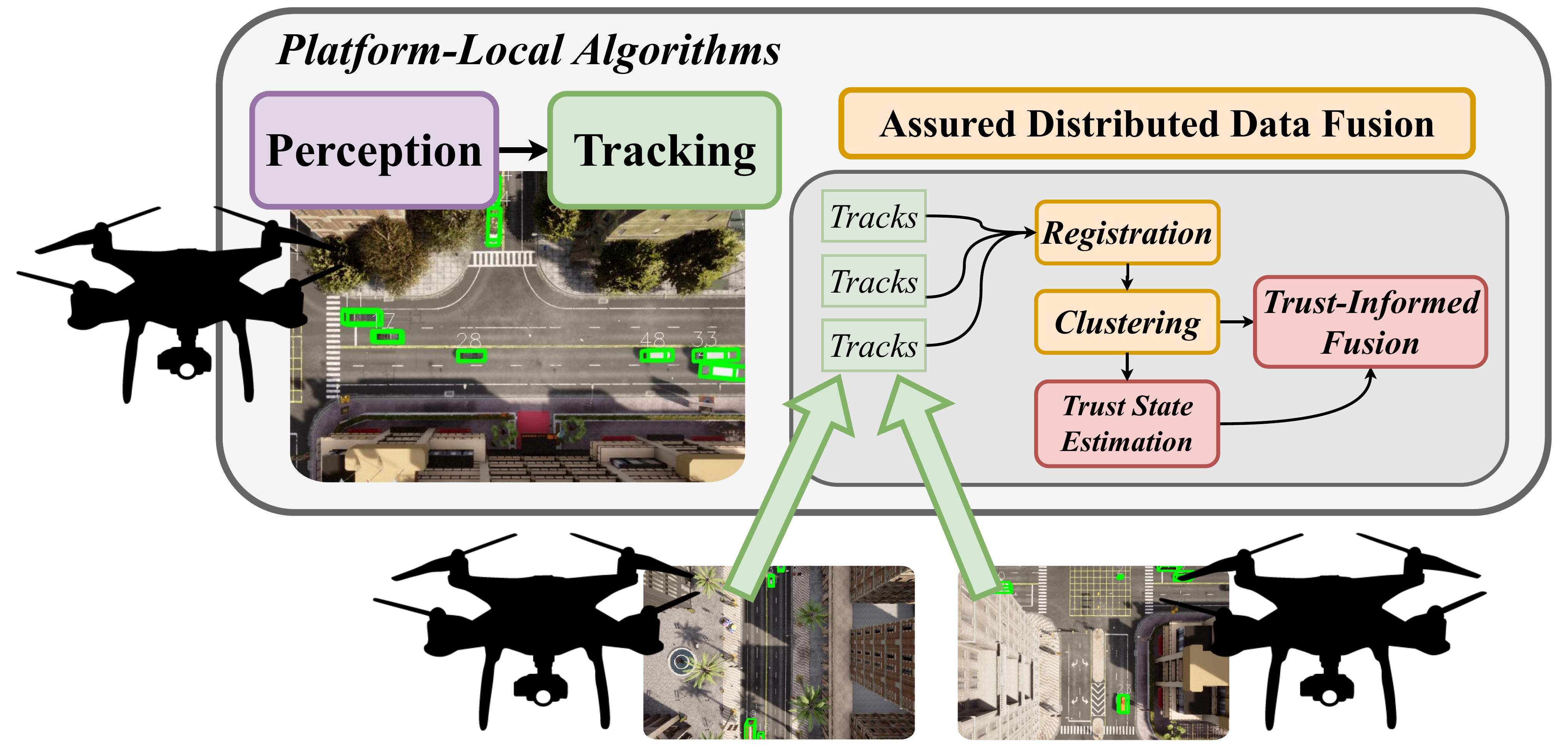}
    \caption{Assured distributed data fusion fuses platform-local object states with tracks from nearby agents. Trust estimation monitors consistency between agent data and weights influence of each agent in trust-informed fusion.}
    \label{fig:assured-ddf}
\end{figure}

We formulate a joint problem of trust estimation and sensor fusion using a hidden Markov model (HMM). In~\eqref{eq:fusion-posterior}, the sensor fusion posterior was presented without trust. By introducing trust, we augment the posterior as 
\[
\Pr(X_{t}, \Tau^c_{t}, \Tau^a_{t} | Z_{1:t}, A_{1:t}),
\]
where the inclusion of trust variables leads to a trust-aware posterior distribution. The principles of conditional probability allow us to decompose this posterior into two subproblems:
\begin{equation}\label{eq:trust-posterior}
\begin{aligned}
    \Pr(&X_{t}, \Tau^c_{t}, \Tau^a_{t} | Z_{1:t}, A_{1:t}) \\
     &= \Pr(\Tau^c_{t}, \Tau^a_{t} | Z_{1:t}, A_{1:t}) \Pr(X_{t} | \Tau^c_{t}, \Tau^a_{t}, Z_{1:t}, A_{1:t}).
\end{aligned}
\end{equation}
This joint problem is addressed via a two-step process of \textbf{(1) estimating agent and track trust} -- i.e., solving $\Pr(\Tau^c_{t}, \Tau^a_{t} | Z_{1:t}, A_{1:t})$, and \textbf{(2) performing trust-informed DDF} -- i.e., computing \\$\Pr(X_{t} | \Tau^c_{t}, \Tau^a_{t}, Z_{1:t}, A_{1:t})$. Figure~\ref{fig:assured-ddf} illustrates the integration of these two steps into the distributed data fusion pipeline for assured~DDF.

\subsection{Distributed Trust Estimation}

In the considered scenarios, agents share navigation and perception data within a distributed ad hoc network. Here, we outline the trust estimation process in this context. To estimate trust, we derive \emph{pseudomeasurements} (PSMs) from the semantic-level information exchanged between nearby agents. These PSMs are then used to update trust estimates for both agents and tracked objects.

\subsubsection{Field of view filtering.} \ Estimating trustedness based on shared situational awareness requires predicting what an agent \emph{should} have observed at any given time. This prediction depends on each agent's field of view (FOV) when comparing tracks between the agents. The FOV changes dynamically with each frame, influenced by the agent's position, sensor characteristics, and object/infrastructure locations. Fortunately, for UAVs, sensors usually have an unoccluded view, unlike ground vehicles that face significant occlusions. Therefore, FOV is mainly determined by sensor characteristics. 

For a gimbaled downward-facing camera with rectangular pixels, the horizontal and vertical angles of the camera’s FOV are
\begin{equation}\label{eq:camera-fov}
    \begin{aligned}
        \Delta \theta = \arctan\left( \frac{n_x}{2 f_x} \right), \quad
        \Delta \phi = \arctan \left( \frac{n_y}{2 f_y} \right),
    \end{aligned}
\end{equation}
where $\Delta \theta$ is the horizontal angle, $\Delta \phi$ the vertical angle of the camera FOV, $n_x$ and $n_y$ are the image dimensions, and $f_x$ and $f_y$ are the focal lengths in pixels. These viewable angles are then used to filter the set of objects down to those the agent was expected to observe.

\subsubsection{PSM functions.} \ Each platform maintains local sensing and performs perception and sensor fusion on its own data. Fusion results are shared between platforms and DDF occurs locally to each platform. Trust estimation at the level of DDF requires first deriving trust measurements from the distributed fusion results to be used to update estimates of trust distributions. 

\begin{figure}
    \centering
    \includegraphics[trim={0 0.5cm 0 0cm},clip,width=\linewidth]{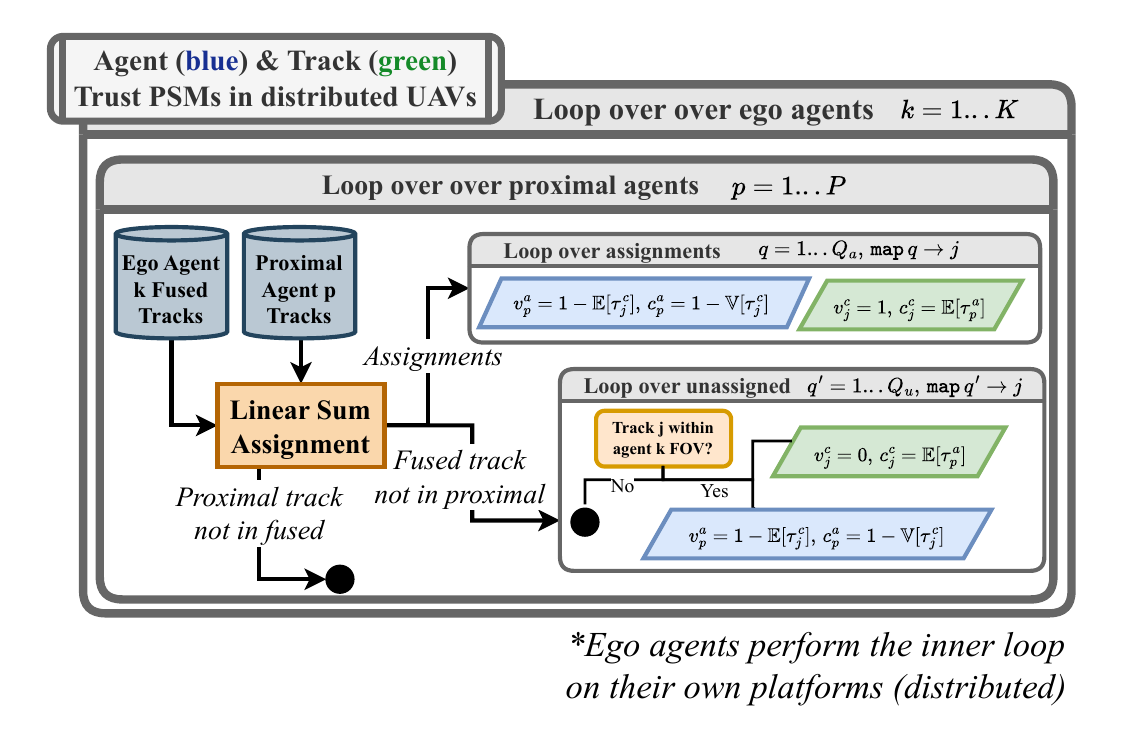}
    \caption{Trust pseudomeasurement (PSM) function compares fused tracks to proximal agent tracks on each of the distributed platforms. Positive associations yield PSMs proportional to agent trust while unassigned tracks are inversely proportional to trustedness. The inner loop occurs on each platform during inter-platform communication.}
    \label{fig:trust-measurements}
\end{figure}

We construct PSM algorithms that map sensor data to the domain of trust on $[0,1]$ as a tuple of value and confidence, $( v_{j,k}, \, c_{j,k})$, providing probabilistic indicators of trustedness. We define a trust PSM function based on assignment proximity between the proximal and ego agent tracks. This measurement generation process is described in Figure~\ref{fig:trust-measurements}. PSMs can then can be incorporated into the Bayesian framework via a Bernoulli likelihood function.

\subsubsection{Leveraging prior information.} \ A strength of Bayesian estimation is in the ability to incorporate prior information, offering an advantage over traditional MBDs. The Beta distribution, $\text{Beta}(\alpha, \beta)$, proposed in~\cite{hallyburton2024bayesian}, is well-suited for trust modeling on the domain \([0,1]\). Moreover, appropriate informative priors reduce reliance on uncompromised agents for situational awareness, if available. Unlike approaches requiring a majority of secure agents (e.g.,~\cite{pajic2014robustness,pajic_tcns17,khazraei_automatica22}), strategic use of secure agents and priors lets trusted agents outweigh uncertain ones, while newly introduced agents can start with neutral priors, limiting influence until trust is~established.

\subsubsection{Iterative trust updates with Gibbs sampling.} \ During measurement generation, we obtain PSMs for agents and tracks. The full trust distribution is a complex multivariate with cross-correlations. This makes it difficult to sample from/estimate the parameters of the joint trust distribution. Luckily, the conditionals
\begin{equation}
\begin{aligned}
    (1)\ &\text{Track trust:} \ \Pr(\Tau^c_{t}\ |\ \Tau^a_{t-1}, Z_{1:t}, A_{1:t}) \\
    (2)\ &\text{Agent trust:} \ \Pr(\Tau^a_{t}\ |\ \Tau^c_{t}, Z_{1:t}, A_{1:t}),
\end{aligned}
\end{equation}
remove the cross-correlations. As such, we choose to let PSMs update parameters of the conditionals and propagate those effects back to the joint posterior. 

This strategy is inspired by Gibbs sampling, a popular technique of sampling a multivariate probability density (PDF). Suppose there is a PDF, $\Pr(x,y)$ that is difficult to sample but whose conditionals, $\Pr(x|y),\, \Pr(y|x)$, are easy to sample. Gibbs sampling is an iterative Markov-chain Monte Carlo method of drawing samplings representative of the joint distribution by sampling solely the conditionals.

\subsubsection{Distribution independence assumption.} \ The full distribution of trust of all agents and all tracks falls on a hypercube with domain $[0,1]^{J\times K}$ for $J$ tracks and $K$ proximal agents. Unfortunately, there is no parameterization for a distribution on this domain without inapplicable constraints (e.g.,~Dirichlet). For efficiency of update, we approximate that the distributions are independent and can be modeled by $JK$ one-dimensional state estimates. We find this assumption converges rapidly in practice. Future works will investigate non-parametric sampling techniques for the general case.

\subsubsection{Updating trust estimates with PSMs.} \ Using a Beta prior with a Bernoulli likelihood function forms a  ``conjugate pair'' and yields a Beta posterior. The parameter update given the prior and the 
generated trust PSMs has a closed form. The derivation of Bayesian updating can be found in e.g.,~\cite{shemyakin2017copula}. With PSMs represented as a tuple of value and confidence, the update to the two-parameter Beta distribution of trust state density at time $t$ will be,
\begin{equation}\label{eq:beta-bernoulli-update}
    \begin{aligned}
        \alpha^c_{j,t} &= \alpha^c_{j,t-1} + \Delta \alpha^c_{j,t},\quad \Delta \alpha^c_{j,t} = \sum_k c_{j,k} v_{j,k} \\
        \beta^c_{j,t} &= \beta^c_{j,t-1} + \Delta \beta^c_{j,t},\quad \Delta \beta^c_{j,t} = \sum_k c_{j,k} \cdot (1-v_{j,k}).
    \end{aligned}
\end{equation}
The resulting $(\alpha^c_{j,t},\beta^c_{j,t})$ are the new Beta parameters for track $j$'s trust. The same process applies for agent trust, $(\alpha^a_{k,t}, \beta^a_{k,t})$. Since the Beta only has two parameters and the update is in closed form, the updates are fast and efficient. The update requires minimal computational overhead and can scale to large numbers of agents. 

\subsubsection{Negatively-Weighted Update.} \ The update in~\eqref{eq:beta-bernoulli-update} weighs positive and negative PSMs equally. This does not match the conventional understanding that trust ought to be built only on large amounts of consistent data but lost on small instances of inconsistency. To reflect this guidance in the trust update, we consider a negatively-weighted update leveraging a tunable weighting parameter, $\omega$, and negativity bias/threshold $\tracknegativitybias$ and $\tracknegativitythreshold$, as captured~here
\begin{equation}\label{eq:beta-bernoulli-update-weighted}
    \begin{aligned}
        \begin{aligned}
            \Delta \alpha^c_{j,t} = \sum_k c_{j,k} &\, v_{j,k} \quad \Delta \beta^c_{j,t} = \sum_k \omega_{j,k} \, c_{j,k} \, (1-v_{j,k})\\
            \omega_{j,k} &= \left.\begin{cases}
                \tracknegativitybias & v_{j,k} < \tracknegativitythreshold \\
                1.0 & \text{otherwise} 
            \end{cases} \right\}.
        \end{aligned}
    \end{aligned}
\end{equation}

\subsection{Trust-Informed Sensor Fusion}

In ISR applications, the primary tasks in multi-target tracking (MTT) through DDF are:
\begin{enumerate}
    \item \textbf{Target existence determination}: This involves distinguishing between real and false tracks.
    \item \textbf{State estimation}: This entails estimating the target's state parameters, such as position and velocity.
\end{enumerate}
Using trust to inform data fusion can foster assured autonomy. Trust-informed DDF integrates trust into MTT tasks via:
\begin{itemize}
    \item \emph{Trust-weighted DDF updates}: These updates apply to state estimation, incorporating trust to refine state estimates.
    \item \emph{Trust-informed track determination}: Trust assesses the authenticity of tracks, influencing existence determination.
\end{itemize}

\subsubsection{Trust-weighted DDF updates.} \ In an ad hoc network, each agent receives information from a subset of proximal agents. Due to the local filtering within each of the agents' pipelines and the unstructured network topology, DDF algorithms presented in Section~\ref{sec:system-model} fuse state and covariance data in the information form using \emph{covariance intersection}. Naively and in the absence of any trust awareness, each data element is either weighed uniformly (naive Bayesian fusion) or is set to minimize some statistic such as the trace of the resulting state covariance~\cite{1994ddfframework}.

To incorporate trustedness for assured DDF, we propose using agents' trust distributions as the weights. Algorithm~\ref{alg:trust-weighted-ci} outlines the trust-informed weighting methodology for augmented covariance intersection. Simply, the expected value of the trust distribution weighs the influence of each agent's update to the state and uncertainty. Moreover, the variance of trust distributions contributes to an aggregate ``fusion confidence'' criterion, $\hat{\zeta}_{CI}$ that is useful for monitoring the general trustedness of fusion outcomes.

\begin{algorithm}[!t]
\caption{N-Fold Uniformly-Trust-Weighted CI}
\label{alg:trust-weighted-ci}
\begin{algorithmic}[1]
\Require Covariance matrices $\Sigma_1$, $...$, $\Sigma_n$, mean estimates $\mu_1$, $...$, $\mu_n$, trust distributions $\tau_1$, $...$, $\tau_n$
\State Initialize weights $\{\omega_i = \expectation[\tau_i], i=1, ..., n \}$
\State Normalize weights $\omega_i = \frac{\omega_i}{\sum_i \omega_i}$
\State Compute trust-weighted fused covariance: 
\[
\hat{\Sigma}_{CI} = \left( \sum_i \omega_i \Sigma_i^{-1} \right)^{-1}
\]
\State Compute trust-weighted fused mean: 
\[
\hat{\mu}_{CI} = \hat{\Sigma}_{CI} \left( \sum_i \omega_i \Sigma_i^{-1} \mu_i \right)
\]
\State Compute trust-weighted fusion confidence: 
\[
\hat{\zeta}_{CI} = \sum_i \omega_i \variance[\tau_i]
\]
\State \Return $\hat{\mu}_{CI}, \hat{\Sigma}_{CI}, \hat{\zeta}_{CI}$
\end{algorithmic}
\end{algorithm}


\subsubsection{Trust-informed track determination.} \ Tracks at the ego agent may represent true objects or be false alarms. Furthermore, in contested environments, tracks aligned with true objects may have partial state estimates that inaccurately capture the true state (e.g.,~correct position state estimate but compromised velocity states). Track trust is used to flag potential false alarms and tracks with potentially compromised state estimates. Any track with a low trust (i.e.,~$\expectation[\tau^c_j] < \thresholdtrackignore$ with $\thresholdtrackignore$ a predetermined threshold) is flagged. 
\section{Multi-Agent Simulation} \label{sec:multi-agent-sim}

To test the distributed trust estimation and data fusion in UAVs, we developed, to our knowledge, the first perception-oriented UAV datasets from the CARLA simulator~\cite{dosovitskiy2017carla} adapting the ground-vehicle dataset pipeline from~\cite{hallyburton2023datasets}. In this section, we discuss the motivation for using CARLA, the dataset requirements, and the evaluation methodology used in this work.

\subsection{CARLA Simulator}
The CARLA simulator provides a platform for capturing physics-based data using realistic sensor models. CARLA’s flexible API allows precise control over a platform’s altitude, orientation, and flight path, enabling the capture of diverse scenes and lighting conditions. Available camera sensors include RGB, depth, and segmentation varieties. CARLA enables the systematic collection of aerial platform imagery in a controlled, reproducible virtual environment, making it a powerful tool for advancing autonomous navigation and perception research.

\subsection{Dataset Generation}
We use the CARLA simulator along with the expanded simulator API from~\cite{hallyburton2023avstack} to construct multi-agent datasets for analysis of baseline and attacked scenes. We place 50 agents in the environment at randomized positions with predefined trajectories. The agents are endowed with local sensing in the form of gimbaled cameras providing ground-facing imagery at 10~Hz data rates. 

The data are labeled with visible ground truth object locations to provide fair evaluation of sensor fusion performance. The data generation pipeline and dataset are released online\footnote{\url{https://cpsl.pratt.duke.edu/research/aerial-dataset}.}~\cite{trust-links}. The dataset pipeline can be modified to suit future experimental needs. For simplicity, we consider predetermined agent trajectories to obtain varied coverage over a large urban environment in CARLA.

\begin{figure}
    \centering
    \begin{subfigure}[t]{0.48\linewidth}
        \centering
        \reflectbox{
            \rotatebox[origin=c]{180}{\includegraphics[cfbox=black 10pt 10pt,width=\linewidth]{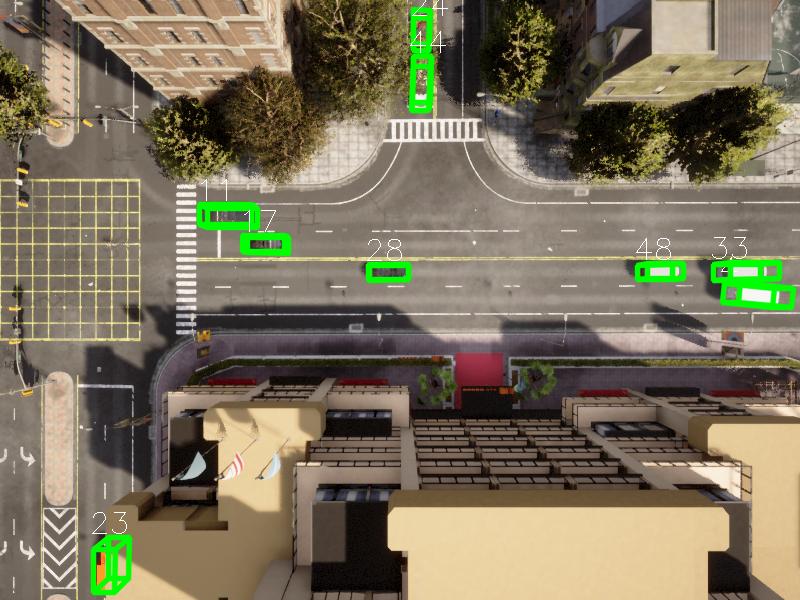}}
        }
        \caption{Agent two view.}
    \end{subfigure}
    \hfill
    \begin{subfigure}[t]{0.48\linewidth}
        \centering
        \includegraphics[cfbox=black 10pt 10pt,width=\linewidth]{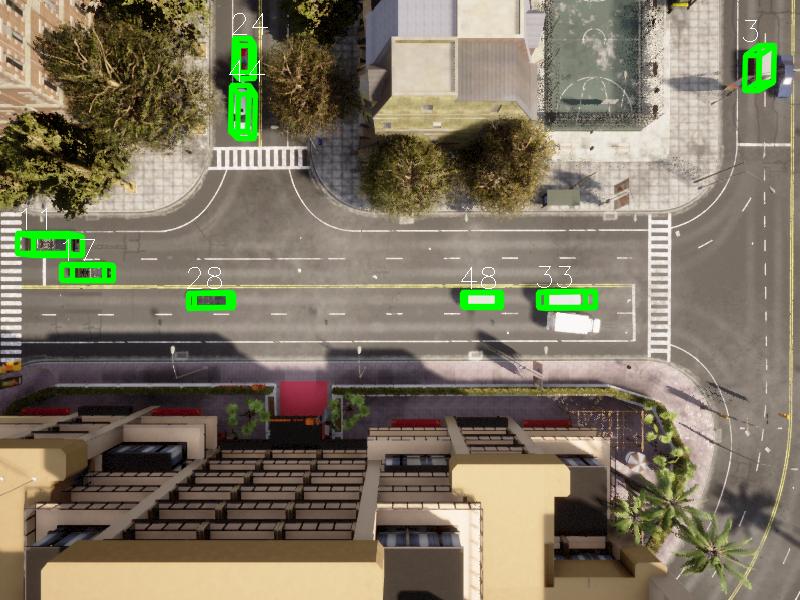}
        \caption{Agent four view.}
    \end{subfigure}
    \caption{Agents two and four observe overlapping regions of the CARLA town allowing for assured distributed sensor fusion through trust estimation and trust-weighted DDF.}
    \label{fig:agent-views}
\end{figure}

\begin{figure}
    \centering
    \includegraphics[trim={1cm 1cm 1cm 1.7cm},clip,cfbox=black 10pt 10pt,width=0.8\linewidth]{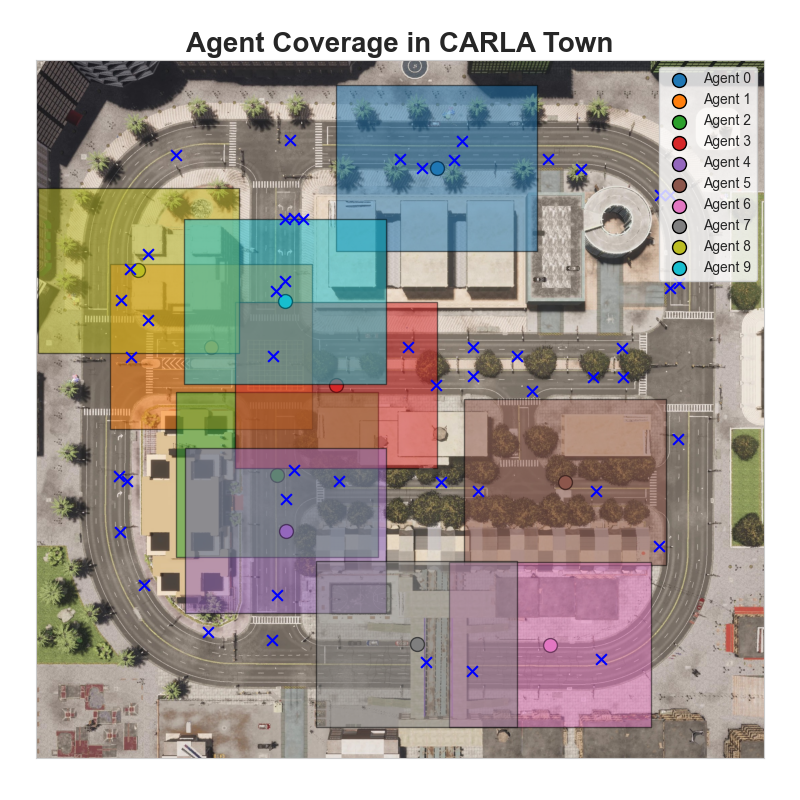}
    \caption{$N=10$ agents perform surveillance over a large region. Agents have partially overlapping fields of view, supporting redundancy of computation for security-aware trust-based methods. In this configuration, Agent 0 is isolated with no inter-agent FOV overlap leading to no trust observability. On the other hand, Agent 3 has overlap with four agents, yielding significant trust confidence.}
    \label{fig:bev-agent-coverage}
\end{figure}
 
\subsection{Evaluations on Datasets}
To perform parametric analysis on the datasets, we employ randomization in post-hoc analysis to create a diverse array of unique trials. To evaluate the effect of number of agents on the sensor fusion and trust estimation performance, we subsample the agents to suit the experiment needs. For example, to study the effect of density of agents on the system performance, we run multiple Monte Carlo trials selecting a subset of agents for each trial to suit the density parameter. Furthermore, to study the effect of communication range on performance, we apply a communication model in postprocessing which allows for greater experiment flexibility.

\subsection{Metrics}

Evaluating distributed ISR requires objective criteria measuring MTT performance. Moreover, trust estimation protocols require dedicated metrics. We employ both industry-standard MTT evaluation metrics and define trust-oriented metrics. Additional details are provided in Appendix~\ref{appendix:metrics}.

\vspace{4pt}
\noindent \textbf{Assignment metrics.} The classical metrics precision, recall, and F1-score (P, R, F1) are functions of the number of true objects, false positives, and false negatives.

\vspace{4pt}
\noindent \textbf{OSPA metric.} The Optimal Sub-Pattern Assignment (OSPA) measures MTT performance~\cite{schuhmacher2008ospa}. OSPA penalizes state estimation error in conjunction with a cardinality penalty for mismatches in number of tracks and truths.

\vspace{4pt}
\noindent \textbf{Track trust accuracy.} Track trust is a PDF on $[0,1]$. The goal of track trust estimation is to have all estimated probability density near 1 for a true object and near 0 for an FP. Track trust accuracy quantifies performance as the area above/below the trust cumulative distribution function (CDF).

\vspace{4pt}
\noindent \textbf{Agent trust accuracy.} Similarly, agent trust is a PDF on $[0,1]$ with trusted agents having probability density near 1 and 0 otherwise.
\section{Assured Distributed Data Fusion Experiments} \label{sec:experiments}

\subsection{Case Study: Secure Sensor Fusion Applied} \label{sec:case-study-prior}

We present a case study demonstrating the effectiveness of our trust estimation and secure sensor fusion pipeline, as shown in Figure~\ref{fig:bev-agent-coverage}. 
For this evaluation, we randomly select a subset of 25 agents from the dataset, each configured to run DDF facilitating information sharing among neighboring agents. 

\begin{figure}
    \centering
    \begin{subfigure}[t]{\linewidth}
        \includegraphics[width=0.92\linewidth]{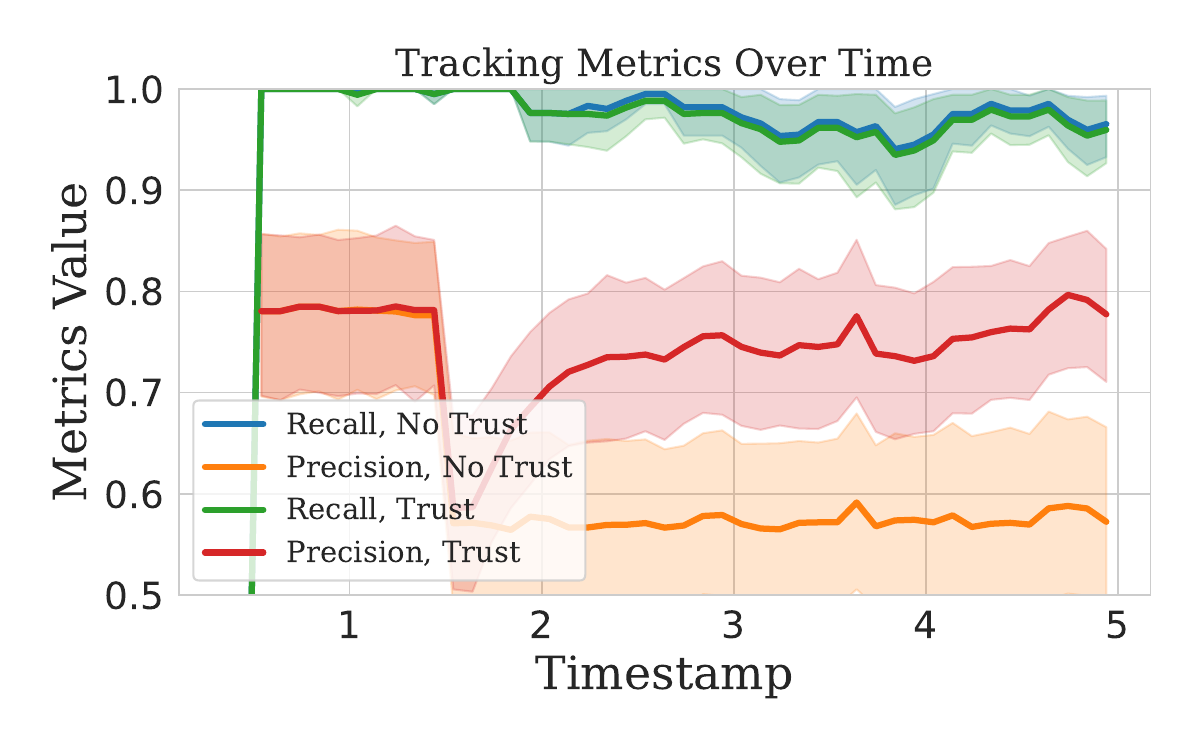}
        \caption{Classical assignment metrics evaluated over case study. Natural false positives exist, meaning baseline precision is imperfect.}
        \label{fig:results-case-metrics-a}
    \end{subfigure}
    \begin{subfigure}[t]{\linewidth}
        \includegraphics[width=0.92\linewidth]{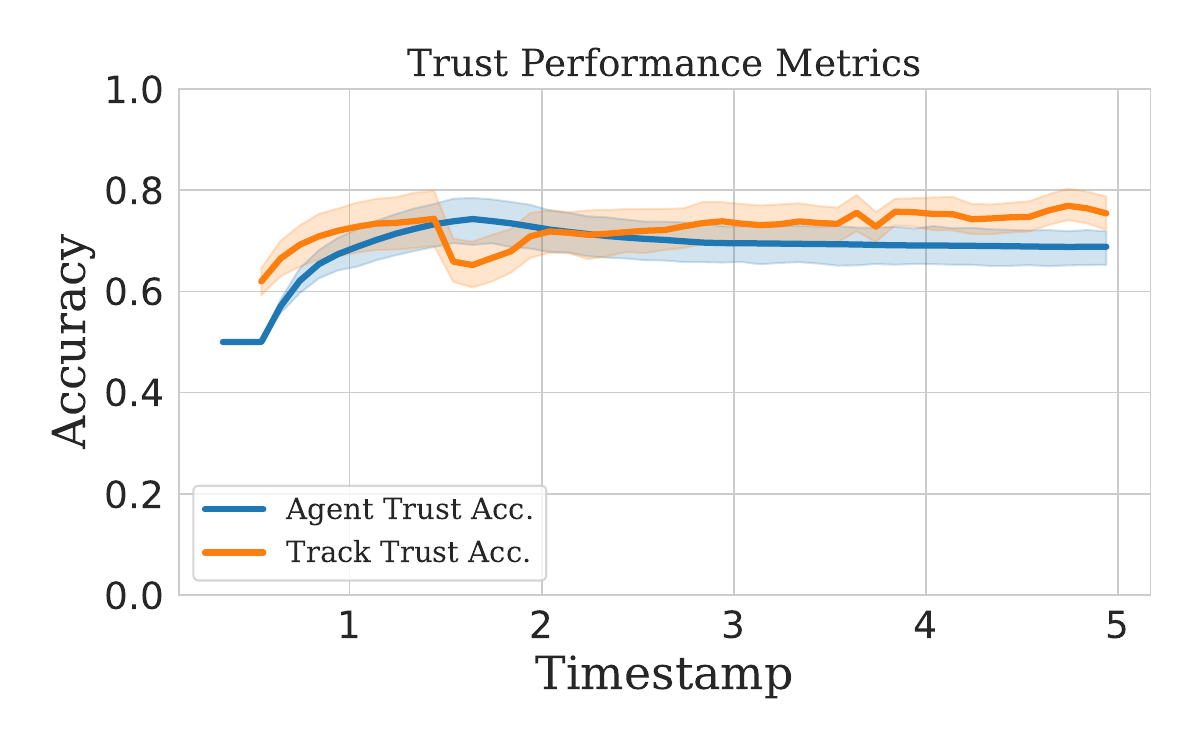}
        \caption{Trust accuracy as the estimated trust distributions' proximity to true agent/target adversarialness. Near $80\%$ indicates distributions are close to targets with uncertainty due to natural sensor noise.}
        \label{fig:results-case-metrics-b}
    \end{subfigure}
    \caption{Attack starts at $t=1.5~s$. Agents suffer natural false positives, meaning baseline precision is not perfect. Uncertainty bounds represent deviations in performance between agents and over multiple trials with randomization. (a) Without trust, precision drops quickly after false positive attack commences. With trust estimation and assured sensor fusion, after initial drop, precision restored to pre-attack baseline. (b) Trust-based metrics continue to improve over time as more information and observability is present. When attack commences, trust performance initially declines while trust estimator converges to new estimated trust states; system is adaptive to changes and does not assume fixed trust.}
    \label{fig:results-case-metrics}
\end{figure}

Following an initial stable period without attacks, a subset of agents is compromised by a false-positive (FP) attacker. In the early stages of the attack, the monitoring capability of agents deteriorates across both pipelines—those with and without trust estimation; performance metrics thus initially drop for all cases. However, after the initial transient window has elapsed, the secure sensor fusion pipeline detects anomalous agents and tracks, subsequently isolating them from the fusion process. Metrics indicate trust-based fusion recovers accurate and robust situational awareness. This action effectively restores the system’s performance to near-baseline levels, as illustrated in Figure~\ref{fig:results-case-metrics-a}. Additionally, the protocol provides accurate estimations of both agent trustworthiness and track reliability, as depicted in Figure~\ref{fig:results-case-metrics-b}.

\begin{figure}
    \centering
    \begin{subfigure}[t]{\linewidth}
        \includegraphics[width=\linewidth]{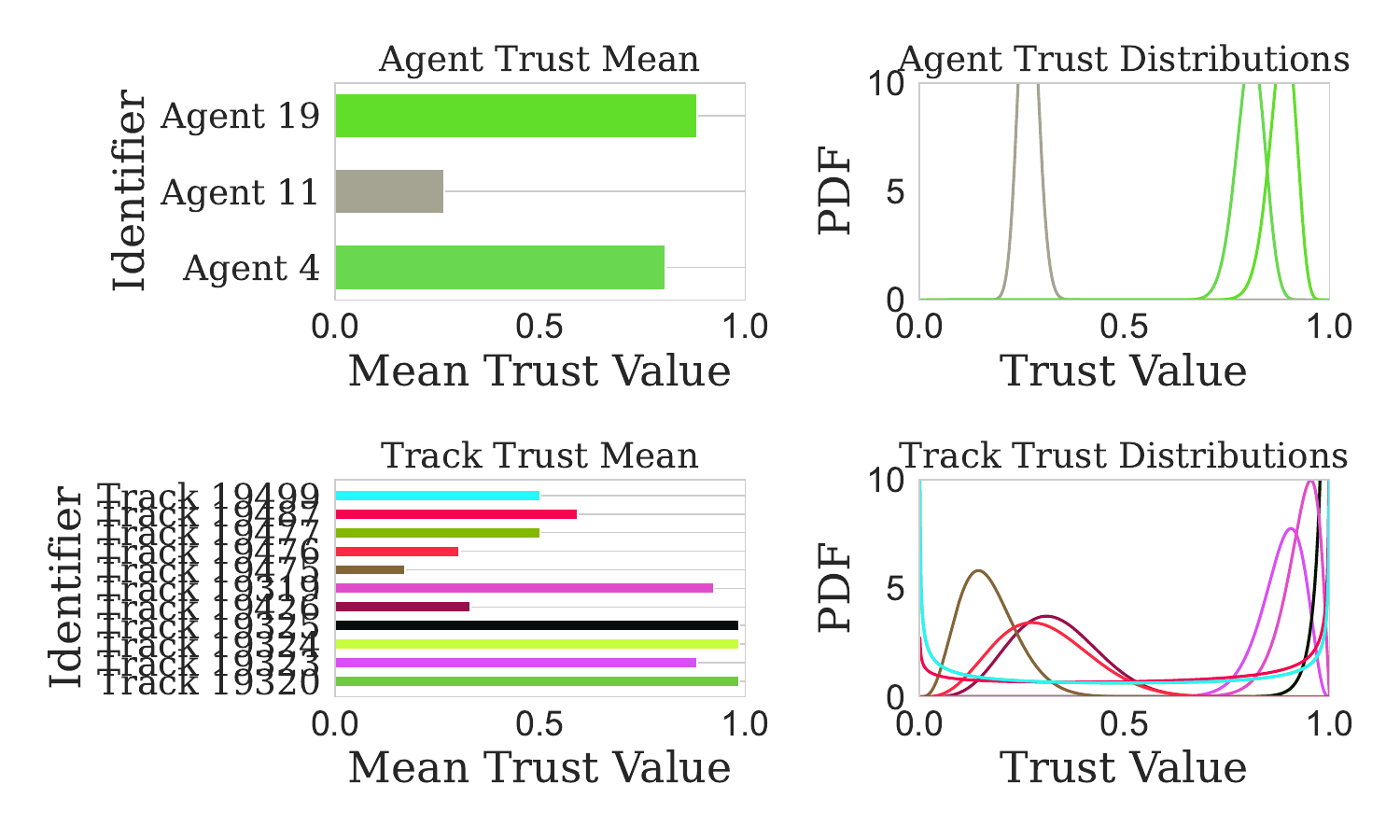}
        \caption{In Agent 19's view, Agent 11 is compromised an provides many false tracks identified with trust estimation.}
        \label{fig:results-case-dist-a}
    \end{subfigure}
    \begin{subfigure}[t]{\linewidth}
        \includegraphics[width=\linewidth]{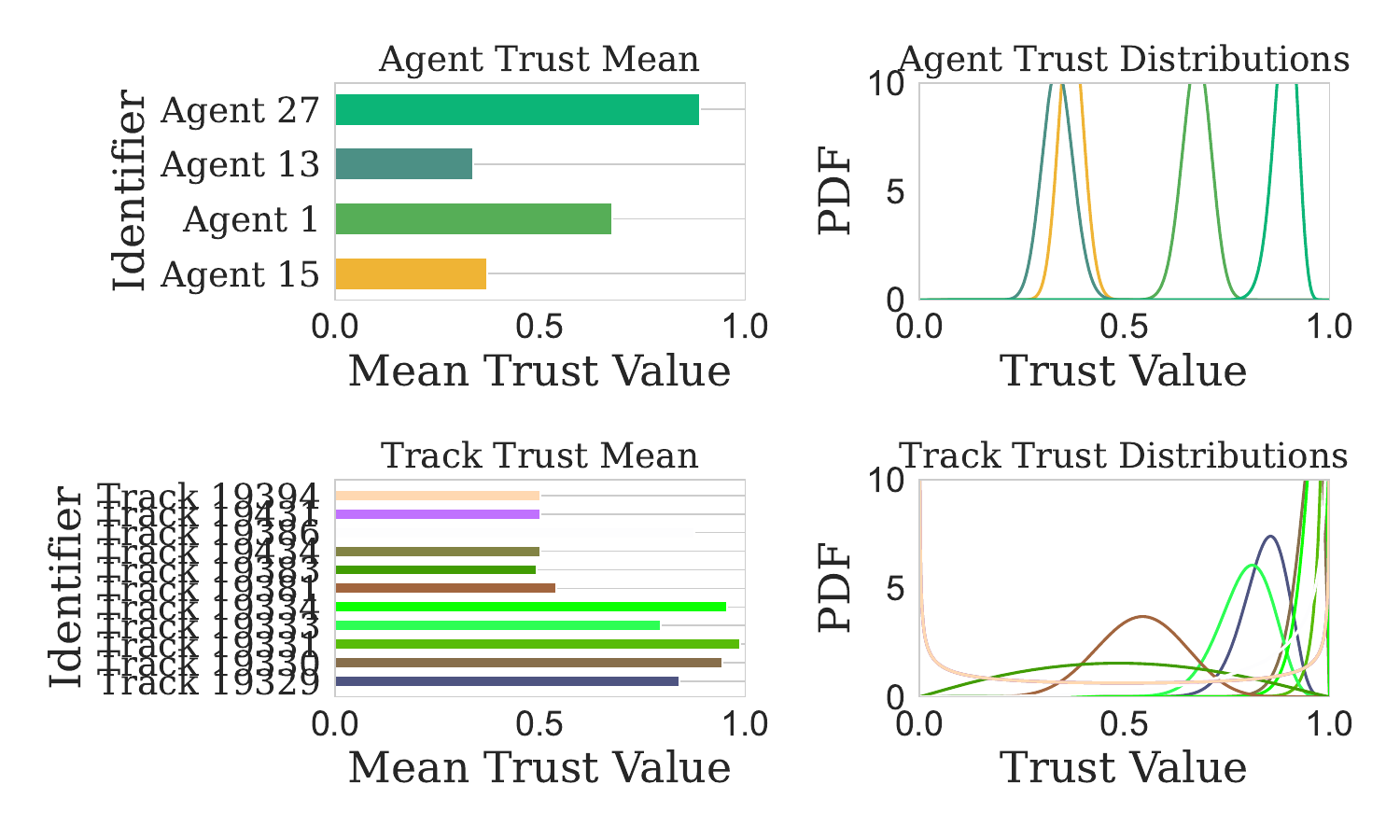}
        \caption{In Agent 27's view, two agents are compromised providing tracks of unknown trustedness.}
        \label{fig:results-case-dist-b}
    \end{subfigure}
    \caption{Trust estimation yields distributions over agent and track trust probability on $[0,1]$ domain. Trust provides useful insight into behavior of agents and data with recursive update based on PSMs generated by comparing local and fused track data.}
    \label{fig:results-case-dist}
\end{figure}

This case study underscores the robustness of the secure sensor fusion approach, which not only mitigates the impact of adversarial agents but also maintains the integrity of trust estimation within the network. The final estimated trust distributions for select agents are shown in Figure~\ref{fig:results-case-dist}.


\subsection{Evaluation: Agent Density} \label{sec:case-study-dense}

The effectiveness of trust estimation in a network of agents hinges on the extent to which agents can share relevant sensor data. In a sparse network, where agents have minimal overlapping FOVs, the potential for redundant sensor fusion is severely limited, restricting the ability to detect adversaries. Conversely, in a densely populated network of aerial agents, high levels of observability can be achieved, significantly enhancing the capacity to identify compromised agents.

In this evaluation, we conduct a parametric study of assured sensor fusion as a function of agent density. Using the CARLA simulator, we generate multiple scenarios with up to 50 aerial agents equipped with wide-angle, downward-facing cameras. Each agent is endowed with local perception and tracking algorithms and operates within the DDF framework, allowing real-time data exchange.

\begin{figure}
    \centering
    \begin{subfigure}[t]{\linewidth}
        \includegraphics[width=0.9\linewidth]{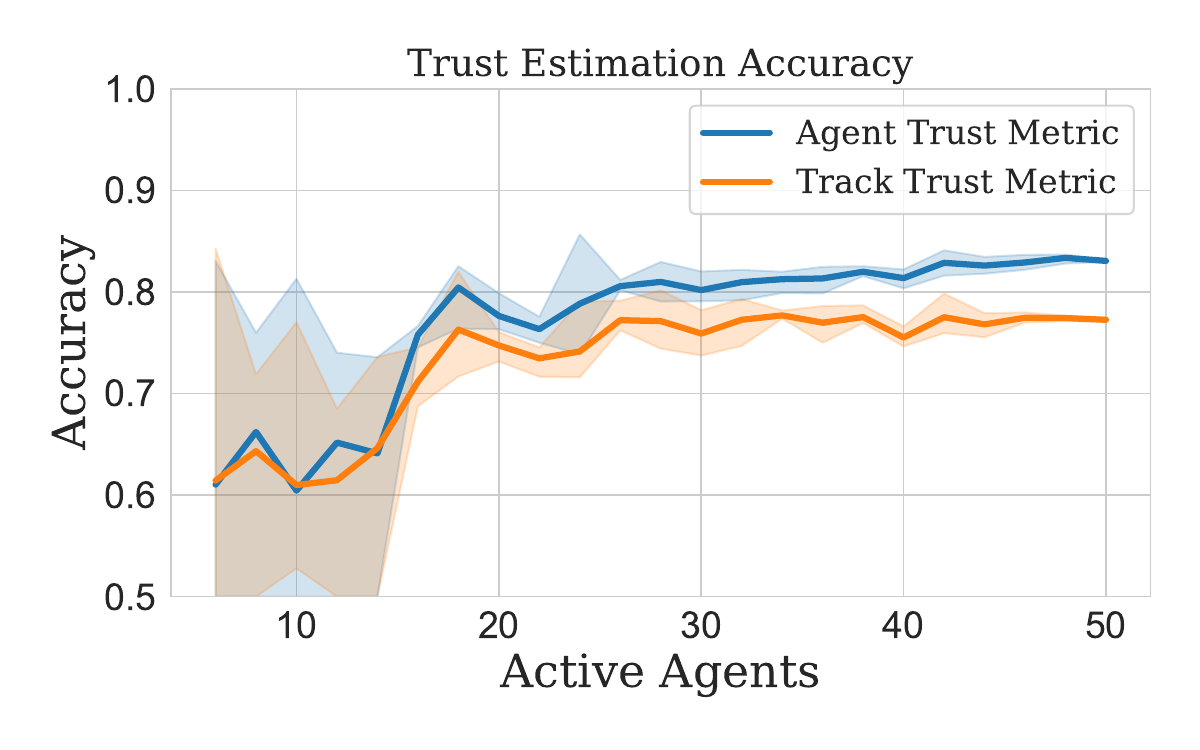}
        \caption{Trust estimation accuracy in benign context when no agents are attacked and all data are consistent improves with higher density.}
        \label{fig:results-density-a}
    \end{subfigure}
    \begin{subfigure}[t]{\linewidth}
        \includegraphics[width=0.9\linewidth]{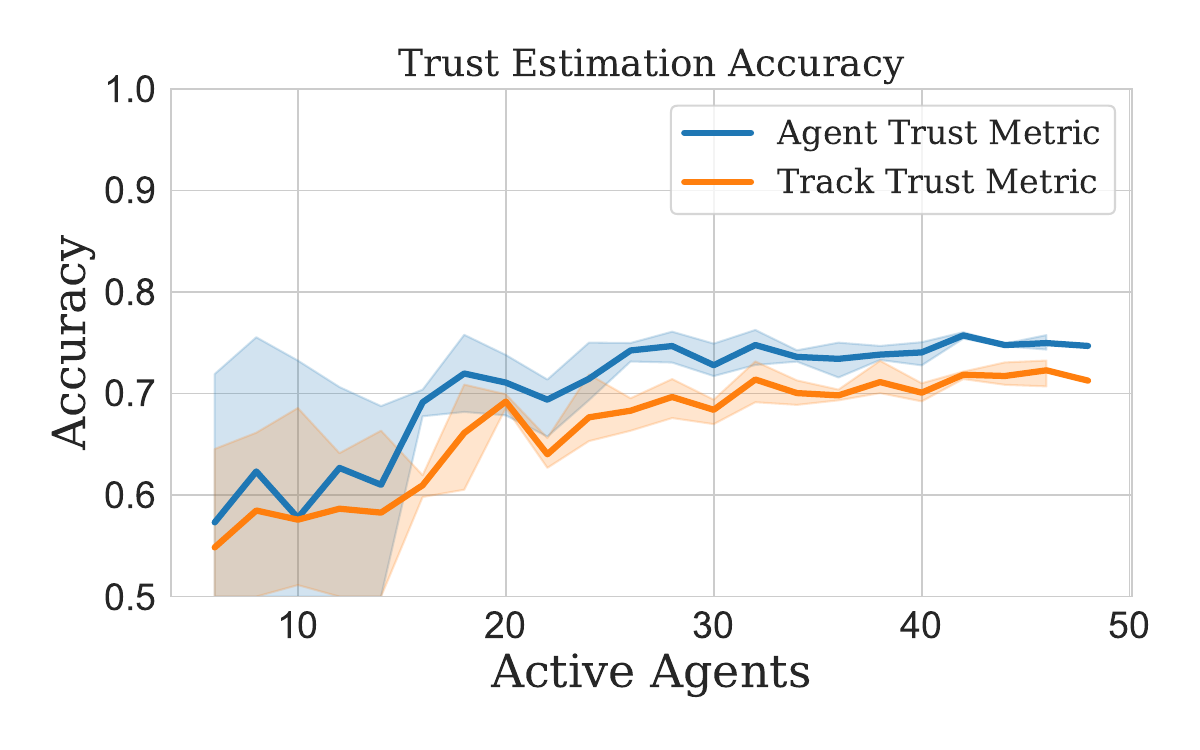}
        \caption{Trust estimation accuracy under mixture of FP and FN attacks on up to 50\% of agents improves with higher density.}
        \label{fig:results-density-b}
    \end{subfigure}
    \caption{Trust estimation accuracy improves significantly with increased agent density in both benign and adversarial environments. In scenarios with fewer agents, accuracy is limited due to reduced observability, highlighting the importance of agent density for reliable trust assessment.}
    \label{fig:results-density}
\end{figure}

To assess the impact of density on fusion reliability, we vary the active agent population from 20\% to 100\% of the total 50 agents, running our assured fusion pipeline across both benign and adversarial conditions. Figure~\ref{fig:results-density} presents the results of these density evaluations. In both benign (Figure~\ref{fig:results-density-a}) and adversarial (Figure~\ref{fig:results-density-b}) environments, the accuracy of trust estimation improves consistently as the network density increases. Note that perfect (1.0) trust accuracy is unattainable due to both unavoidable FOV gaps where some objects are only observable by a single agent and also trust distributions subject to naturally noisy sensor measurements. 

Overall, the results demonstrate a clear trend: denser agent networks significantly enhance the robustness of trust estimation against potential adversarial threats.

\subsection{Evaluation: Attacker Capability} \label{sec:results-capability}

\begin{figure}
    \centering
    \begin{subfigure}[t]{\linewidth}
        \includegraphics[width=0.9\linewidth]{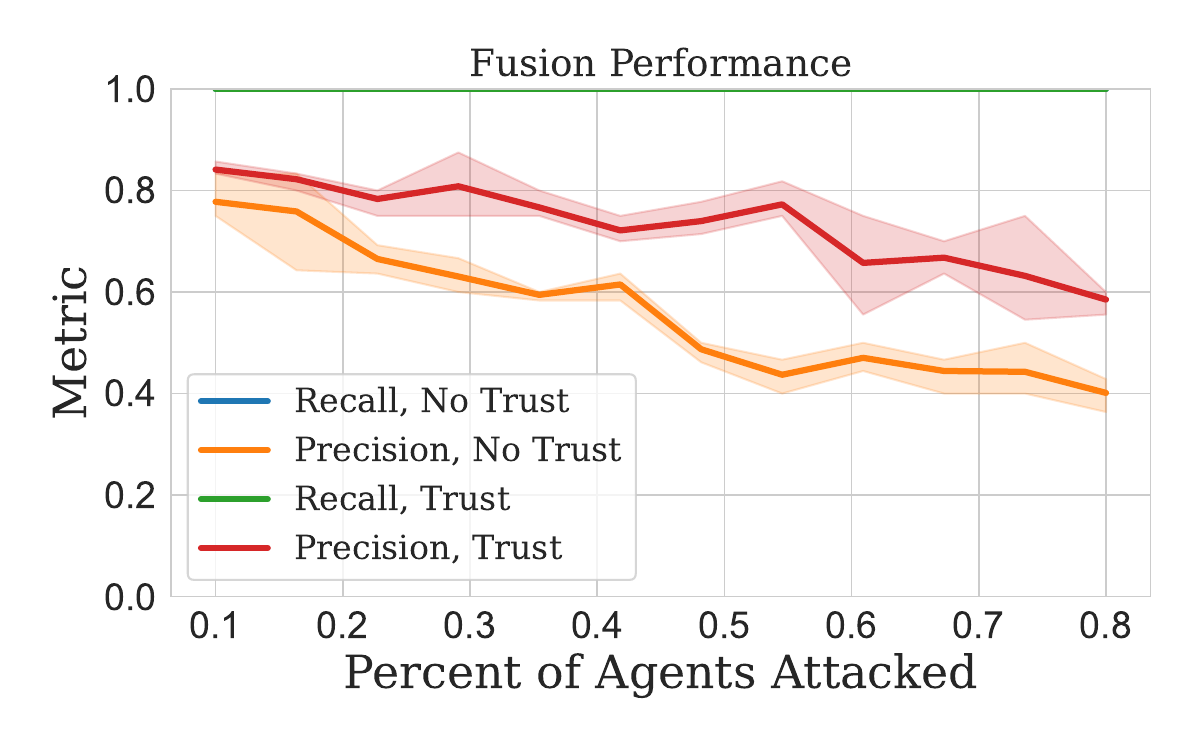}
        \caption{Trust estimation improves tracking performance even when large numbers fractions of agents are attacked.}
        \label{fig:results-capab-a}
    \end{subfigure}
    \begin{subfigure}[t]{\linewidth}
        \includegraphics[width=0.9\linewidth]{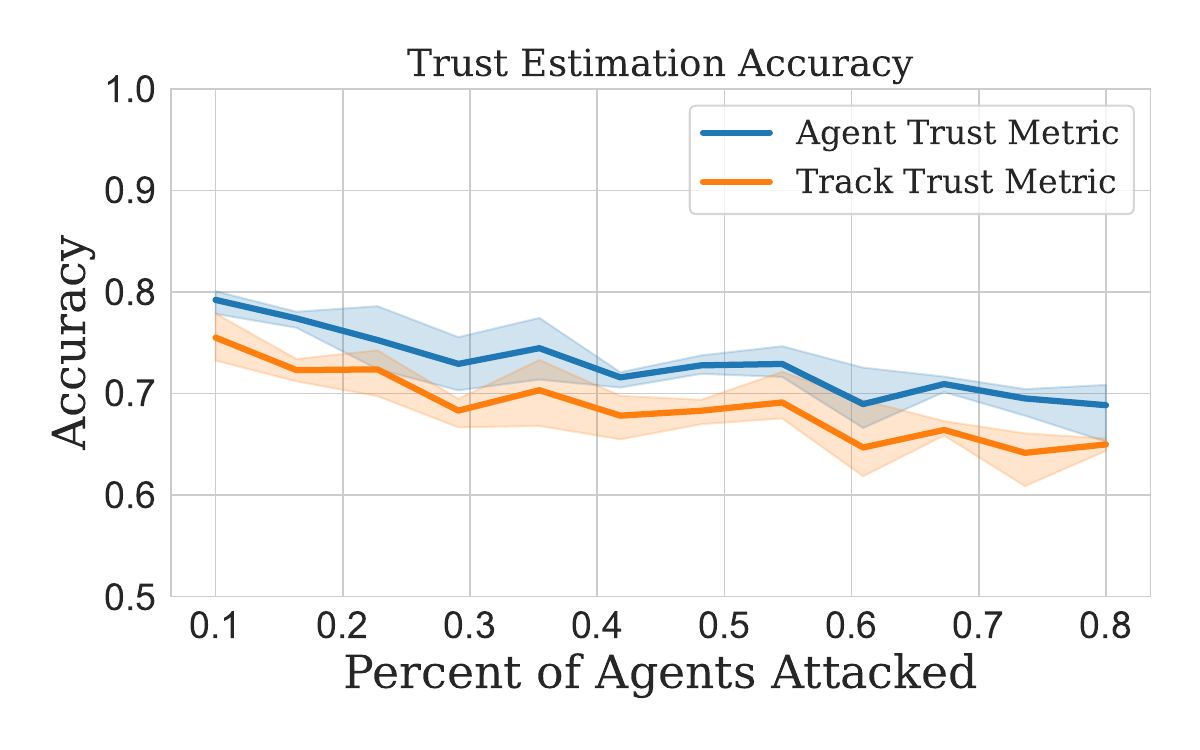}
        \caption{Trust algorithm still able to accurately estimate agent and track trust when many agents compromised.}
        \label{fig:results-capab-b}
    \end{subfigure}
    \caption{(a) Track fusion performance declines as the number of attacked agents increases. However, integrating trust estimation into the sensor fusion pipeline significantly enhances algorithm resilience, enabling more robust operation in contested environments. (b) Trust estimation accuracy decreases with a higher proportion of attacked agents, as consistency checks between compromised agents become increasingly challenging.}
    \label{fig:results-capab}
\end{figure}

To evaluate the impact of the number of attacked agents on sensor fusion performance, we conduct a parametric experiment by systematically increasing the attacker’s capability. Adversarial agents are progressively introduced, injecting false positive detections into the sensor fusion process. Classical assignment-based metrics, such as precision and recall, are captured at the sensor fusion level. Results indicate that as the percentage of attacked agents increases, sensor fusion performance steadily declines, as shown in Figure~\ref{fig:results-capab-a}. Additionally, the system's ability to estimate agent trustworthiness also diminishes with a higher number of attacks, as illustrated in Figure~\ref{fig:results-capab-b}. Despite these challenges, trust-based sensor fusion consistently shows significant improvements in precision and recall, even under adversarial conditions. 

It is important to note that the adversary model is applied \emph{randomly} across the agent’s entire FOV. In regions where false positives are injected without overlap from other agents, these false positives cannot be detected using existing methods. This modeling choice implies that some degree of performance degradation is unavoidable by construction due to the randomness of overlap of FOVs. These findings emphasize the robustness of trust-based sensor fusion while highlighting the challenges posed by adversaries in contested environments.

\section{Discussion}

\subsection{Future Research Directions}

\subsubsection{Segmentation-based consistency.} \ Current methods for deriving trust PSMs rely on comparing estimated states of objects detected from camera images across agents. In aerial vehicles with camera-based sensing, trust PSMs could instead leverage pixel-wise semantic segmentation over the \emph{entire} image. This form of consistency evaluation would enable detection of manipulations to image sources, enhancing security and robustness. However, such an approach would require increased bandwidth to transmit the larger data volumes associated with segmentation maps.

\subsubsection{Trust-informed mission planning.} \ By continuously estimating the trustworthiness of agents and tracked objects, mission planners can optimize agents' planned trajectories to maximize observability and reliability. For instance, highly trusted agents could be tasked with investigating agents or tracks of unknown trustworthiness. Integrating trust estimation into mission planning ultimately bolsters the resilience and robustness of multi-agent operations, particularly in high-stakes or contested environments.

\subsubsection{Prediction-informed trust estimation.} \ The PSM algorithm relies on correspondence of tracked objects in overlapping regions of agents' FOVs. A larger number of objects in the FOV increases observability into agents’ trustworthiness. The number of corresponding objects can be expanded by considering temporal dynamics: specifically, tracks within an ego agent's FOV could be compared to tracks from other agents' FOVs at previous time points that were \emph{predicted} to appear in the ego's FOV. Utilizing prediction would provide a more extensive basis for consistency evaluation.

\subsection{Limitations of Trust Estimation}

\subsubsection{Dependency on agent coverage.} \ The effectiveness of trust estimation depends on redundancy in the ad hoc network. When agent coverage is sparse or agents have narrow FOVs, this redundancy and therefore the trust observability is limited. Section~\ref{sec:experiments} presented a parametric analysis of agent density, illustrating degradation in model performance under sparse coverage.

\subsubsection{Inaccurate prior information.} \ Providing inaccurate prior information on agent trustedness in the Bayesian estimation framework can significantly compromise the ability of the system to detect faults, particularly if the agent density is sparse.
\section{Conclusion}

This work addresses the critical need for assured multi-agent distributed data fusion for ISR in ad hoc UAV networks. We proposed a trust-based assured sensor fusion framework, integrating trust estimation through a Beta distribution-based HMM approach. This framework enables UAVs to collaboratively achieve situational awareness while mitigating the risks posed by adversarial agents and naturally occurring sensor errors. Using newly constructed multi-agent UAV datasets in CARLA, we evaluated the robustness of our approach, demonstrating its effectiveness in enhancing system resilience. Our results underscore the value of trust-based methodologies for assured autonomy in contested environments, establishing a foundation for future research in secure, trust-informed sensor fusion.

\section{Repeatability Package}

The instructions for repeating the experiments for this work can be found on the project repository, \url{https://github.com/cpsl-research/iccps-2025-multi-agent-rep}. Each of the experiments are documented the the \texttt{experiments} folder, and running the experiments will generate all numerical results in this document. The README online details all system requirements and dependencies. We have included for convenience a Docker file at \url{https://hub.docker.com/repository/docker/roshambo919/iccps25/general}. Full instructions can be found in the repository README. 

\begin{acks}
This work is sponsored in part by the ONR under agreement N00014-23-1-2206, AFOSR under the award number FA9550-19-1-0169, and by the NSF under the NAIAD Award 2332744 as well as the National AI Institute for Edge Computing Leveraging Next Generation Wireless Networks, Grant CNS-2112562.
\end{acks}

\bibliographystyle{ACM-Reference-Format}
\bibliography{references}

\appendix
\section{Metrics} \label{appendix:metrics}

\subsection{Assignment Metrics} \label{appendix:metrics-assignment}

Assignment-based metrics describe the performance of perception and tracking. A bipartite matching between e.g., detections and ground truth objects yields false positive, false negative, and true positive matches. Statistics over the assignment outcome including precision, recall, and the F1-score as the harmonic mean between precision and recall are used to summarize the performance of an algorithm. These metrics are defined as

\begin{equation} \label{eq:assignment-metrics}
    \begin{aligned}
        \textbf{Precision:} \quad & \frac{\#\text{TP}}{\#\text{TP} + \#\text{FP}}\\
        \textbf{Recall:} \quad & \frac{\#\text{TP}}{\#\text{TP} + \#\text{FN}} \\
        \textbf{F1 Score:} \quad & \frac{2 \ \text{Precision} \cdot \text{Recall}}{\text{Precision} + \text{Recall}}.
    \end{aligned}
\end{equation}
An adversary adding false positives will affect precision while an adversary making false negative attacks will compromise recall.

\subsection{OSPA Metrics} \label{appendix:metrics-ospa}
The derivation of the Optimal SubPattern Assignment (OSPA) derives from~\cite{schuhmacher2008ospa}. OSPA is a mixing of state estimation error, a penalty for unassigned tracks/truths, and a penalty for different numbers of tracks/truths. Given a bipartite assignment between the set of tracks and the set of truths, $\mathcal{A} = \{a_{i,\text{track}},\, a_{i,\text{truth}}\}_{i=1}^{n_a}$, with $n_a = |\mathcal{A}|$ the number of assignments, the \texttt{cost} of an assignment as $\texttt{dist}(a_{i,\text{track}},\,a_{i,\text{truth}})$, and the unassigned tracks/truths as $\Bar{\mathcal{A}}$, OSPA is
\begin{equation} \label{eq:ospa-metric}
    \begin{aligned}
        \textbf{OSPA} \coloneqq \frac{1}{n} \left( \sum_{i}^{n_a} \mathcal{A}_i.\texttt{cost}() + c \, |\mathcal{\Bar{\mathcal{A}}}| \right)^p + c^p (n - m)
    \end{aligned}
\end{equation}
with $c$ and $p$ user-defined weighting parameters, $n$ and $m$ the smaller and larger of the number of tracks and truths, respectively, and $|\mathcal{\Bar{\mathcal{A}}}|$ the smaller of the number of unassigned tracks and unassigned truths. Lower OSPA implies lower cost/penalty and is desirable. 

\subsection{Trust Estimation Accuracy} \label{appendix:metrics-trust}
We build trust-oriented accuracy metrics for both track and agent trust. Both use the distance between a reference PDF $f_r(\tau)$, (i.e., the estimated trust) and a target PDF, $f_t(\tau)$, (i.e., the true trust). The distance is the area between the reference and target cumulative distribution functions (CDFs), $F_r(\tau),\, F_t(\tau)$, i.e.,
\begin{equation}
    \begin{aligned}
        D_{\tau} &= \int_0^1 |F_t(\tau) - F_r(\tau)| d\tau.
    \end{aligned}
\end{equation}
For a true trust that is binary (completely trusted or distrusted), the target CDF is either $F_t(\tau) = 1.0$ for false/distrusted or $F_t(\tau) = 0.0 \ \texttt{if} \ \tau < 1 \ \texttt{else} \ 1.0$. Thus, for a false/distrusted target, $F_t(\tau)$ majorizes $F_r(\tau)$ while for a true/trusted target, $F_t(\tau)$ minorizes $F_r(\tau)$. This allows us to remove the absolute value in different target cases to simplify the integration as
\begin{equation}
    \begin{aligned}
        D_{\tau} &= \begin{cases}
            1 - \int_0^1 F_r(\tau) d\tau & \text{target is false/distrusted} \\
            \int_0^1 F_r(\tau) d\tau & \text{target is true/trusted}
        \end{cases}
    \end{aligned}
\end{equation}
Moreover, with integration by parts on any CDF $F_X(x)$ whose PDF $f(x)$ has support on $[a,b]$ (i.e.,~$F_X(a)=0,\, F_X(b)=1$),
\begin{equation}
    \begin{aligned}
        \int_a^b F_X(x) dx &= \left[x F_X(x)\right]_a^b - \int_a^b F'_X(x) dx \\
        &= (bF_X(b) - aF_X(a)) - \int_a^b x f(x) dx \\
        &= b - \expectation[f(x)]
    \end{aligned}
\end{equation}
As a result, the distance simplifies to
\begin{equation} \label{eq:trust-distance}
    \begin{aligned}
        D_{\tau} &= \begin{cases}
            \expectation[f_r(\tau)] & \text{target is false/distrusted} \\
            1 - \expectation[f_r(\tau)] & \text{target is true/trusted.}
        \end{cases}
    \end{aligned}
\end{equation}
The remaining consideration for track and agent trust distances ($D_{\tau}^{\text{track}},\, D_{\tau}^{\text{agent}})$ is to determine the target distribution class (i.e., false/distrusted vs. true/trusted). 

\subsubsection{Track trust metric, $D_{\tau}^{\text{track}}$.} \ For each track state with its associated reference trust distribution, the target label is \texttt{true} if  the track-to-truth assignment yielded a positive assignment for the track in question and \texttt{false} otherwise.

\subsubsection{Agent trust metric, $D_{\tau}^{\text{agent}}$.} \ For each agent with its associated reference agent trust distribution, we maintain an oracle set of the agents that are attacked and assign a target of \texttt{distrusted} to agents in that set and \texttt{trusted} to all other agents. An alternative is to consider trustworthiness according to a maintained F1-score.

\end{document}